\documentclass[prl,twocolumn,showpacs,preprintnumbers,amsmath,amssymb]{revtex4}
\usepackage{dcolumn}
\usepackage{bm}
\usepackage{graphicx}
\usepackage{bm}
\begin{document}
\title{Formation of finite antiferromagnetic clusters and the effect of electronic phase separation in   Pr${_{0.5}}$Ca${_{0.5}}$Mn${_{0.975}}$Al${_{0.025}}$O${_3}$}
\author{Sunil Nair and A. Banerjee}
\affiliation{Inter University Consortium for D.A.E. Faclilites,\\ University Campus, Khandwa Road, Indore, 452 017, INDIA.}
\date{\today}
\begin{abstract}
We report the first experimental evidence of a magnetic phase arising due to the thermal blocking of antiferromagnetic clusters in the weakened charge and orbital ordered system Pr${_{0.5}}$Ca${_{0.5}}$Mn${_{0.975}}$Al${_{0.025}}$O${_3}$. The third order susceptibility ($\chi{_3}$) is used to differentiate this transition from a spin or cluster glass like freezing mechanism. These clusters are found to be mesoscopic and robust to electronic phase separation which only enriches the antiphase domain walls with holes at the cost of the bulk, without changing the size of these clusters. This implies that Al substitution provides sufficient disorder to quench the length scales of the striped phases. 
\end{abstract}
\maketitle 
Cuprates and manganites belong to the class of highly correlated electron systems, where the kinetic and potential energies share comparable scales resulting in a fascinating array of electronic and magnetic properties. The parent compounds in both these class of materials are antiferromagnetic (AFM) insulators and the physical properties evolve as a function of hole doping. Current consensus is that the AFM host has the tendency to expel away holes, an electronic phase separation (EPS) mechanism leading to regions with varying hole concentrations\cite{dag}. When coupled with other competing interactions like the Coulombic repulsion, this leads to various self organised structures. In practical systems, this could take the form of an antiphase domain wall\cite{tranquada}, along which the holes are concentrated and across which the antiferromagnetic order changes sign. An interesting experimental manifestation of this charge segregation is the formation of the \emph{cluster spin glass} phase in cuprates which has been a subject of intense scrutiny for many years and has been detected using a variety of measurements\cite{chou,julien}. This phase is postulated to appear when AFM clusters encircled by the antiphase domain walls freeze, provided there is sufficient disorder in the system to prevent long range ordering. However, there has been no conclusive evidence of such a state arising purely out of EPS in the mixed valent manganites . Here, hole concentrations of close to 50$\%$ are most relevant for studies of this nature since Coulombic repulsive forces are known to stabilise the formation of a charge and orbital ordered (COO) AFM ground state\cite{tokura}. Thus it is not surprising that a variety of relaxation effects possibly arising due to phase competition have been seen in these compositions\cite{dag1}.

 In manganites, the disorder necessary for prohibiting the long range order can be easily introduced in the form of impurities in the Mn-O-Mn network. The nature of the dopant has a crucial bearing on the magnetism, as impurities with partially filled d bands (like Cr or Co) could participate in the broadening of the e${_g}$ bandwidth\cite{kim1}. The bandwidth (W) of the host system is equally important as shown by the fact that substitution of Cr in the large W system La${_{0.46}}$Sr${_{0.54}}$MnO${_3}$ is seen to induce a rentrant spin glass phase\cite{dho}, whereas when substituted in the narrow W system Nd${_{0.5}}$Ca${_{0.5}}$MnO${_3}$ the formation of a relaxor ferromagnet is speculated\cite{kimura}. Co substitution in Pr${_{0.5}}$Ca${_{0.5}}$MnO${_3}$ is reported to form an inhomogeneous magnetic state which also exhibits sharp magnetisation steps as a function of applied field\cite{mahend}. Substitution studies where the impurity has a non magnetic configuration (d${^0}$ or d${^{10}}$) would be particularly advantageous, since it would be expected to introduce random disorder in the COO state without introducing any magnetic interactions of its own. Ionic considerations also need to be taken into account, considering the fact that lattice distortions play a crucial role in stabilising the physical properties of these systems.

 In an attempt to observe the evolution of magnetic phases in a COO system with quenched non magnetic impurities, we have undertaken a detailed study of the compound Pr${_{0.5}}$Ca${_{0.5}}$Mn${_{0.975}}$Al${_{0.025}}$O${_3}$  using dc magnetization (DCM) and linear and nonlinear ac susceptibility (ac$\chi$). Since large applied fields can mask the intrinsic signatures of inhomogenously magnetised systems, we have concentrated on \emph{low field} measurements to discern the magnetic ground state. Non magnetic Al was selected on ionic considerations and polycrystalline samples were prepared using the solid state ceramic route, the details of which have been reported earlier\cite{sunil}. In this letter, we report the first conclusive observation of a thermal \emph{blocking} of AFM spin clusters in a weakened COO system, contrary to the cluster spin glass \emph{freezing} observed in the cuprates. Moreover, the size of these clusters is found to be robust within the experimental time scales  against EPS, which only enriches the domain walls with holes at the cost of the antiferromagnetism of the bulk.
 
The parent Pr${_{0.5}}$Ca${_{0.5}}$MnO${_3}$, with a narrow W is known to be a prototype COO system, and a robust CO is seen to set in at T$\approx$240K\cite{jirak}. This real space ordering of charge carriers, accompanied by an alternate ordering of the d${_{3x{^2}-r{^2}}}$/d${_{3y{^2}-r{^2}}}$ Mn${^{3+}}$ orbitals in the \emph{ab} plane is manifested in bulk magnetic measurements in the form of a rapid fall of the susceptibility at T${_{CO}}$. This COO is known to enhance antiferromagnetic fluctuations and long range AFM order sets in at T${_N}$$\approx$175K. Interestingly, the correlation lengths for the Orbital Order (OO) is seen to be shorter than that of the CO, indicating the presence of an orbital (antiphase) domain state\cite{zim}. Al substitution in the Mn site of Pr${_{0.5}}$Ca${_{0.5}}$MnO${_3}$ introduces random disorder in the COO state, as a result of which the strength of the CO reduces, as is shown in the inset of Fig. 1.

Dilution of the magnetic lattice using non magnetic Al causes the long range AFM transition temperature (T${_N}$) to drop to $\approx$ 50K, as is shown in Fig. 1. Interestingly, an additional feature is observed at lower temperatures as evidenced by a cusp in the Zero Field Cooled (ZFC) DCM at T${_G}$ $\approx$ 26K. Moreover, a strong irreversibility in DCM is observed, as indicated by the large bifurcation in the ZFC and field cooled (FC) magnetisation measurements. This history dependence is a generic feature of disordered systems, and is seen in a variety of systems like spin glasses(SG), cluster glasses (CG) and  superparamagnets (SPM). It is interesting to note that unlike in a prototype SG\cite{mydosh}, here the irreversibility is seen to commence at temperatures much higher than T${_G}$. Also, below T${_G}$ the FC magnetization is seen to increase monotonically with decreasing temperature, where as in a spin glass, the FC magnetisation would be expected to remain almost independent of temperature. 

However these observations are by no means conclusive; since it is known that both SG and SPM with large dipolar interactions can show a bifurcation in FC and ZFC much above T${_G}$ and SPM's with a narrow volume distribution can exhibit a temperature independent FC below T${_G}$ similar to that of a prototype spin glass. Coupled with the fact that experimental signatures like a large history dependence in DCM, frequency dependence in ac$\chi$ and time dependence in remenant magnetisation are exhibited by all of the above mentioned metastable systems makes identification between them a non trivial exercise. Needless to say, the physics involved in these systems are totally different, since \emph{freezing} is a co-operative phenomenon where spins ( or clusters) are frustrated due to random competing ferro and antiferro interactions, where as \emph{blocking} is purely dynamic phenomenon arising out of the competition between the thermal and the magnetic (volume and anisotropy) energies of the magnetic clusters.

We have circumvented this problem by using the third ordered susceptibility($\chi {_3}$)to determine the nature of the observed low temperature phase. $\chi{_3}$ has long been used as a direct probe of the divergence of the Edward-Anderson order parameter, signifying the onset of a spin glass transition\cite{binder}. It is well known that theoretically, $\chi {_3}$ is expected to have a negative divergence in the limits H $\rightarrow$ 0, and T $\rightarrow$ T${_G}$, where H denotes the magnetic field\cite{suzuki}. Though SPM's also show a negative cusp in $\chi {_3}$ analogous to that exhibited by SG's; the crucial difference is that unlike for a SG, here $\chi {_3}$ is not \emph{critical} as a function of H or T. This intrinsic difference can be used to accurately differentiate a \emph{freezing} from a \emph{blocking} phenomenon\cite{ashna}. This can be vividly demonstrated by plotting the peak value of $\chi {_3}$ as a function of H as is shown in Fig. 2, where the peak value of $\chi {_3}$ is plotted as a function of H for the sample Pr${_{0.5}}$Ca${_{0.5}}$Mn${_{0.975}}$Al${_{0.025}}$O${_3}$. Here, $\chi {_3}$ is clearly seen to saturate in the limit H$\rightarrow$ 0, indicating that this low temperature phase which we have observed, occurs due to the \emph{blocking} of the magnetic entities and is not a co-operative freezing phenomenon. 

The total magnetisation (M) of a system of non interacting (single domain) SPM particles can be given as 
$M = n\left\langle \mu \right\rangle L \left(\left\langle \mu \right\rangle H/k{_B}T\right)$;
where n is the number of particles per unit volume, $\left\langle \mu \right\rangle$ is the average magnetic moment of a single magnetic entity, k${_B}$ is the Boltzmann's constant and L(x) is the Langevin's function. The linear and non linear susceptibilities above the blocking temperature can then be calculated to be 
$\chi{_1} = n\left\langle \mu \right\rangle/3k{_B}T$, and 
$\chi{_3} = - \left(n\left\langle \mu \right\rangle/45\right)\left(\left\langle \mu \right\rangle/k{_B}T\right)^{3}$.
Thus the $\chi{_3}$ is negative and proportional to T${^{-3}}$\cite{wolf}. The inset of Fig. 2 shows this T${^{-3}}$ dependence for the real part of the measured $\chi {_3}$ for this sample, further substantiating the fact that the feature observed in our magnetic measurement arises due to the blocking of magnetic entities.

An estimation of $ \left\langle \mu \right\rangle$ can be made using the ratio $\chi{_3}$/$\chi{_1}$ as is obvious from the above equations. Using fits to $\chi{_3}$ and $\chi{_1}$, the value of $ \left\langle \mu \right\rangle$ determined is of the order of  10${^4}$ Bohr magneton. Here, it is important to note that a prototype SPM will have a large value of $ \left\langle \mu \right\rangle$ due to the fact that it is made up of large number of spins alligned in a single domain. However the effective $\left\langle \mu \right\rangle$ in our case should have been miniscule, as here the clusters have AFM order within them. This value of $\left\langle \mu \right\rangle$ can arise partly due to the presence of uncompensated (free) spins on the surface of these entities. The existence of a finite dipolar interaction term between these surface spins could also explain the experimental observation of the fact that the onset of irreversibility in DCM   measurements is seen much above T${_G}$. Here it is interesting to note that in our case, the AF clusters are blocked, where as in the cuprates the AF clusters are known to be frozen. We believe that this contrasting behaviour occurs due to the rather unique phase diagram exhibited by the narrow bandwidth Pr${_{1-x}}$Ca${_{x}}$MnO${_3}$ compounds, as it is well known that a ferromagnetic (FM) state is not known to occur anywhere in the vicinity of half doping\cite{tom}. Hence, inspite of EPS causing an appreciable spatial distribution of hole concentrations, there is no FM interaction which would facilitate a co-operative freezing mechanism. However, in the cuprates the presence of a finite FM exchange coupling between the Cu spins is well known\cite{aha}, which would promote magnetic frustration. This absence of a FM component is also substantiated by the fact that throughout the temperature range of our measurements, the second order susceptibility ($\chi{_2}$) which arises due to the presence of a symmetry-breaking field\cite{rang} was seen to be absent.

The following two important experimental observations, we believe have significant bearing on the nature of EPS in Pr${_{0.5}}$Ca${_{0.5}}$MnO${_3}$ in particular and manganites in general:(i)T${_G}$ as measured in ac $\chi$ and DCM does not vary, inspite of the fact that the time scales of these probes are vastly different. This would imply that the clusters are reasonably large, and our preliminary estimates show that they are of the order of 500 nm.(ii) Fig. 3 and 4 show the large hysteretic behaviour of ac $\chi$ as a function of thermal cycling. It is to be noted that the direction as well as the range of thermal cycling is different for Fig. 3 and 4. However, there is no observable change in the T${_G}$, indicating that the cluster volume does not change . Thus the size of these clusters is concluded to be robust against EPS. Fig.4 indicates that EPS reduces the hole concentration of the bulk in the cooling run, causing a corresponding reduction in T${_N}$; an observation which is in agreement with both the phase diagram of Pr${_{1-x}}$Ca${_x}$MnO${_3}$, as well as our current understanding of EPS. This reduction of holes in the bulk would enhance its concentration at the domain walls without any observable change in the size of the clusters. This is evident from our observation that a T${^{-1}}$ fit to $\chi{_1}$ measured above T${_G}$ in the heating and cooling cycles indicate that the factor $n\left\langle \mu \right\rangle$ is larger in the cooling run by a factor of 1.3. Moreover,  the magnetic loss as measured by $\chi{_1}{^{I}}$ is seen to be enhanced in the cooling run, presumably arising as a consequence of a larger inter cluster interactions across the hole enriched \emph{antiphase} domain wall. The thermal hysteresis in $\chi{_1}$ is seen to extend to the region T$>$T${_{CO}}$, clearly indicating the presence of superheated CO clusters. This is in agreement with studies indicating the existance of a new scale T${^*}$ where these correlations would form (or die out)\cite{burgy}.

All our experimental observations can be qualitatively explained on the basis of a model incorporating the effects of clusters formed as a result of the destabilisation of the pristine COO state. Al substitution in the COO Pr${_{0.5}}$Ca${_{0.5}}$MnO${_3}$ introduces random impurities in the Mn-O-Mn lattice and reduces the range of the CO interaction, resulting in the formation of (weakened) CO clusters. On cooling, an AFM order is stabilised \emph{within} these clusters; and below T${_G}$ they become thermally blocked with respect to the time scales of our experimental probes. The important point to be noted in all our measurements is that though the magnitude of $\chi{_1}$ changes during thermal cycling, the blocking temperature (T${_G}$) \emph{does not change}. This indicates that there are no appreciable changes in the size of the AFM clusters and that the change in susceptibility arises due to the change in the density of spins at the domain walls; this change being a function of the thermal/magnetic history of the system. A preliminary finite size scaling analysis substantiates this argument, indicating that the AFM correlation length($\xi$) is limited to a maximum value of the order of the cluster size\cite{sunil1}. This would imply that Al substitution in the Mn-O-Mn lattice provides pinning centers, which probably results in the quenching of the length scales of the stripe phases\cite{hassel}. This spatial inhomogeniety in hole concentrations can account for the observation of CE and pseudo-CE phases as detected using neutron diffraction on a similar system\cite{yaicle}. A spin flip transition as a function of applied dc field can also explain the multiple magnetisation steps seen in these systems\cite{mahend}, as the critical field required for triggering a metamagnetic transition would be different for the two phases. 

In summary, we present a new scenario of EPS in manganites where random non magnetic substitution in a COO matrix leads to robust mesoscopic AFM clusters, where the EPS only changes the hole concentration of the bulk of the cluster with respect to its domain walls without invoking any FM interaction. This conclusion has significant impact on the EPS mechanism in transition metal oxides in the presence of disorder and calls for a proper identification of low temperature metastable features observed in these systems.

\begin{figure}
\caption{Irreversibility in DCM as measured using ZFC and FC magnetisation runs in the sample  Pr${_{0.5}}$Ca${_{0.5}}$Mn${_{0.975}}$Al${_{0.025}}$O${_3}$ at a measuring field of 100 Oe. The inset shows the real part of $\chi{_1}$ as measured at a field of 12 Oe and frequency of 133 Hz for the samples Pr${_{0.5}}$Ca${_{0.5}}$Mn${_{0.975}}$Al${_{0.025}}$O${_3}$ and  Pr${_{0.5}}$Ca${_{0.5}}$MnO${_3}$. The reduction of the CO strength with Al substitution is clearly seen.}
\caption{The peak value of $\chi{_3}$ plotted as a function of the applied ac field H, clearly indicating that $\chi{_3}$ is not \emph{critical} with H. Inset shows the temperature variation of $\chi{_3}$ across T${_G}$ for Pr${_{0.5}}$Ca${_{0.5}}$Mn${_{0.975}}$Al${_{0.025}}$O${_3}$. The straight line is a T${^{-3}}$ fit to the data obtained at a field of 12.5 Oe at an exciting frequency of 133 Hz. }
\caption{The real part of $\chi{_1}$ as measured in the cooling and heating runs at an exciting field and frequency of 12.5 Oe and 133 Hz respectively for Pr${_{0.5}}$Ca${_{0.5}}$Mn${_{0.975}}$Al${_{0.025}}$O${_3}$. The star shows the beginning of the temperature cycle, with the arrow indicating the direction. The inset clearly shows that T${_G}$ does not vary with this thermal cycle. }
\caption{The upper panel shows the real part of ac $\chi$ ($\chi{_1}{^R}$) as measured at a field of 2.5 Oe and a frequency of 733 Hz for the sample Pr${_{0.5}}$Ca${_{0.5}}$Mn${_{0.975}}$Al${_{0.025}}$O${_3}$ in heating and cooling cycles. The lower panel shows the imaginary part of ac $\chi$ ($\chi{_1}{^I}$) as measured during the same measurement run. The star indicates the beginning of the thermal cycle with the arrow indicating the direction.}
\end{figure}
\end{document}